# Spectrally Limited Periodic Waveforms for HF OTHR Applications

Yuri Abramovich, *Life Fellow, IEEE,* Dan Dickey, *Member, IEEE,* and Victor Abramovich, *Member, IEEE*

*Abstract* - **The problem of a constant modulus (CM) continuous wave (CW) waveform design with the thumb-tack ambiguity function that meets the NTIA RSEC requirements is addressed. The ad-hoc and alternating projection techniques are proposed to modify the spectrum of a prototype waveform to meet the NTIA RSEC requirements, retaining the "thumb-tack" property of the original waveform ambiguity function. Periodic binary frequency shift keying (BFSK), Costas FSK, and noise-like waveforms are modified to meet CM and NTIA RSEC requirements. Introduced examples demonstrate the spectrum modification consequences and the proposed technique's efficiency for generating the CM spectrum-controlled waveforms with the thumb-tack ambiguity function.**

## I. INTRODUCTION

Most of the Western High-Frequency Over-the-Horizon Radars (HF OTHR), such as the Australian JORN [1, 2] or the US ROTHR [3, 4], operate with the periodic Linear Frequency Modulated Continuous Wave (LFM CW) waveforms.

Among other remarkable properties, these waveforms for high enough compression gain (time-swept bandwidth product) have a very close to rectangular spectrum shape. This property allows for these waveforms to fit the NTIA RSEC Criterion C mask with a very insignificant extension of the occupied bandwidth (OBD) over the deviation (modulation, swept) bandwidth (DBD). Recall that this RSEC mask defines the OBD at the level of -40 dB down from the spectrum maximum and requires further spectrum roll-off at the rate of -20 dB over a decade up to the level of -60 dB [5].

Appropriate for most of the conventional targets, LFM CW waveform application in HF OTHR for modern, very fast (hypersonic) highly maneuvering targets is associated with several serious problems.

Indeed, the repetition frequency in these radars is limited by the "range folding" phenomenon caused by ionospheric propagation and does not exceed ~100 Hz, irrespective of the maximal Doppler frequency of the target of interest.

Specific properties of the "long-ridged" LFM-CW waveform ambiguity function lead to the strong ambiguity of the range and Doppler frequency estimates for targets moving with a radial velocity that exceeds the repetition frequency.

The resolution of this ambiguity becomes a tracker functionality with all the associated problems for multiple highly maneuvering targets.

For this reason, the CW waveforms with the "thumb-tack" ambiguity function should be considered an alternative to the LFM CW, despite considerable complications in the digital signal processing of these waveforms.

Unfortunately, most of the known phase-keying waveforms, such as BPSK (M-sequences) or their Binary Frequency Keying (BFSK) analogs [6, 7], can meet the NTIA RSEC mask requirement only for OBD that significantly exceeds the deviation (modulation) bandwidth (DBD). This is because the instantaneous phase or even frequency change requires a very broad bandwidth for implementation. This problem was recognized in [6], where the authors proposed the Binary Frequency Shift Keying (BFSK) waveform that may be implemented as a phase-continuous periodic waveform with the ambiguity function that mimics the ambiguity function properties of the prototype BPSK waveform.

The introduced in [6] comparison of the BPSK prototype and the BFSK waveforms spectra demonstrated much faster decay of the BFSK spectrum compared with the BPSK one.

Despite the improvement in spectrum shape properties of the BFSK waveform, even for the frequency offset that exceeds the deviation bandwidth by three (!) times ($9/4\tau_0$ vs. $3/4\tau_0$, $\tau_0$ is the chip duration), the BFSK spectrum still does not reach the mask level of -40 dB required by the NTIA RSEC.

The problem of the much broader emission bandwidth compared with the deviation bandwidth for the HF OTH radars is more severe than just compliance with the regulations. Since the bandwidth of the radar waveform is not regulated, compliance may be achieved by claiming a broader (than modulation) bandwidth of a given waveform.

Yet, in a highly congested HF band, the OTHR Frequency Management System has to find an appropriately broad bandwidth unoccupied by other users within a limited HF frequency sub-band that supports ionospheric propagation to the area of interest [8].

Under these conditions, the need for a waveform with the occupied bandwidth (OBD) as close as possible to the deviation bandwidth (DBD), meeting the NTIA RSEC mask requirements, becomes essential indeed.

Note that the recent trend in mobile communication on spectrum sharing with microwave radars makes this problem important for microwave radars as well [10, 11].

Correspondingly, Section II of this paper introduces the prototype waveforms with the "thumb-tack" ambiguity function, considered for spectrum modification.

Section III introduces the ad-hoc approach of spectrum modification developed for FSK waveforms and illustrates its efficiency for the Costas prototype waveform modification [9].

Section IV describes our "analytic" alternating projection technique for spectrum shaping of an arbitrary prototype waveform.

In Sections IV A, B, and C, we analyze the efficiency of the alternate projection technique applied to the Costas waveform, binary FSK prototype waveforms, and the special case when a realization of the broadband white Gaussian noise is used as a prototype to generate a constant modulus waveform with the spectrum meeting the NTIA RSEC mask requirements. This special case may be treated as the new methodology for generating CM CW waveforms with the "thumb-tack" ambiguity function for practical application in HF OTH radars.

In Section V, we conclude our paper.

In Part II of this series, we analyze mismatch processing methodology and efficiency when implemented in Direct Digital Receivers (DDRx) of the modern HF OTHR for mitigation of the "thumb-tack" ambiguity function sidelobes for important radar scenarios.

## II. PROTOTYPE WAVEFORMS SELECTION

The phase-continuous requirement for the periodic CW waveform excludes most phase-keying waveforms from consideration, leaving frequency-modulated waveforms with the thumb-tack ambiguity function as the most viable candidates for being selected as the prototypes.

For this reason, apart from the mentioned above BFSK waveforms introduced in [6], we also considered Costas frequency shift keying (FSK) waveforms [9] as the original waveforms for spectrum modification.

If the ambiguity function properties of these constant modulus FSK waveforms are considered appropriate, the only property that prevents their practical implementation is the spectrum shape of these waveforms.

Let us analyze their spectrum shapes. In Fig. 1 and Fig. 2, we introduce the spectra of the input (Fig. 1) and output (Fig. 2) signal of the radar High Power Amplifier (HPA) for the BFSK waveform.

This waveform was generated using the methodology introduced in [6], with the M = 511-element BPSK M-sequence as the prototype.

According to this methodology, the following parameters of the N = 511 BFSK waveform were selected:

- waveform repetition frequency, $WRF = 48.9237\ Hz$
- elementary chip duration $\tau_0 = 1/(WRF \times 511) = 40.0\ \mu s$
- frequency shift $\Delta f = 1/2\tau_0 = 12.5\ kHz$
- modulation (deviation) bandwidth $BDB = 2\Delta f = 25.0\ kHz$

From Fig. 1, it follows that the occupied bandwidth (defined according to NTIA recommendations at the level of -40 dB down from the spectrum maximum) is three times wider than the deviation bandwidth of 25.550 kHz, observed at the level of -20 dB.

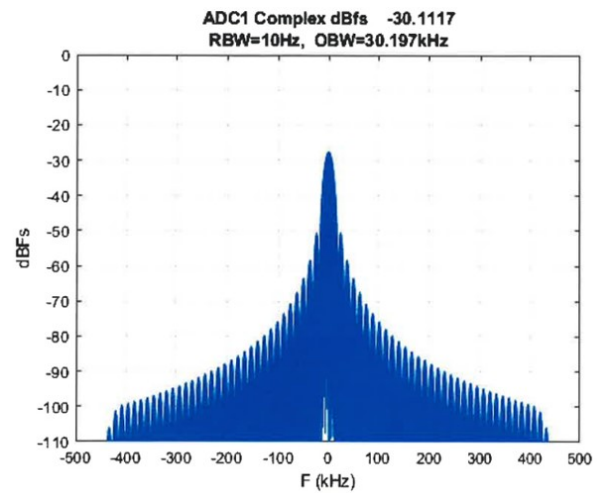

Fig. 1. BFSK signal: Transmitter input.

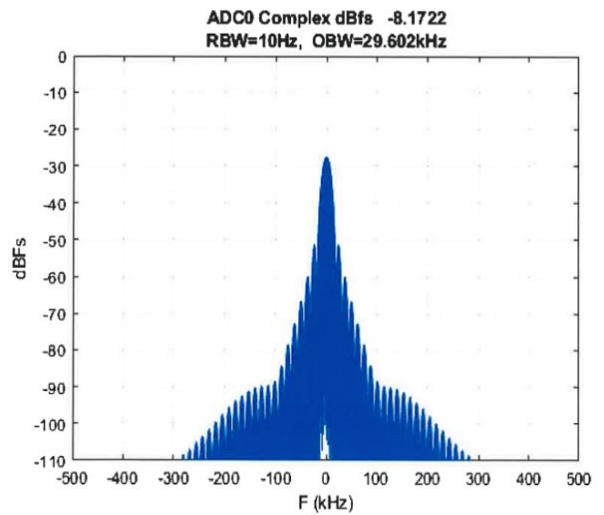

Fig. 2. BFSK signal: Transmitter output.

One can also observe the difference between the HPA input (waveform generation output) and the HPA output spectra below -60 dB levels. This is due to the DSP band-limiting filter within the HPA network.

The HPA DSP-caused spectrum truncation led to insignificant amplitude modulation within the range (0.9997 ÷ 1.0003) (Fig. 3).

The selection of the BFSK waveform as a prototype has also been justified by the remarkable property of the ideal waveform Periodic Autocorrelation Function (pACF), which is very important for HF OTHR applications.

In these applications, the waveform is used as the intra-period modulation of a periodic constant modulus (CM) waveform train. Therefore, the Periodic Auto-Ambiguity Function (pAAF) and the Periodic Autocorrelation Function (pACF), in particular, define the performance of the HF OTH radar.

For example, the range sidelobes of pACF define the dynamic range of the jointly detectable targets moving with the same radial velocity and angle of arrival when the conventional matched waveform processing is being used.

Similar to the BPSK M-sequence prototype, the sidelobe level of the ideal periodic ACF is equal to $-20\log N$, which for $N = 511$ is equal to -54.17 dB. This is much better than the sidelobes of pACF for the LFM CW waveform.

Yet, these properties are of the ideal BFSK waveform, with an unlimited spectrum, as seen in Fig. 1.

The impact of the spectrum modification and, especially, the much more aggressive one required for HF OTHR applications needs to be investigated in detail. This analysis is conducted in Section IV.

The second Multiple Frequency Shift Keying (MFSK) waveform we selected for our analysis was the Costas waveform [9]. Within the repetition period (WRF = 50 Hz), we slotted in 20 "chips" with a frequency offset of 1 kHz so that the modulation (deviation) bandwidth of this waveform is 20 kHz. Therefore, this waveform's compression gain (400) is a bit smaller than for the BFSK waveform above.

In Fig. 4 and Fig. 5, we introduce the input (ideal waveform) and HPA output waveform sampled at a 1 MHz sampling rate.

Once again, spectrum modification provided by the DSP HPA network below the spectrum level of -60 dB could be observed.

Note that the particular M = 20 Costas sequence was not explicitly optimized. Therefore, this waveform's periodic ambiguity function (Fig. 6 and Fig. 7) is not a truly thumb-tack, unlike the ambiguity function of the BFSK waveform (Fig. 8 and Fig. 9) with a very low sidelobe level of pACF and homogeneous pAAF.

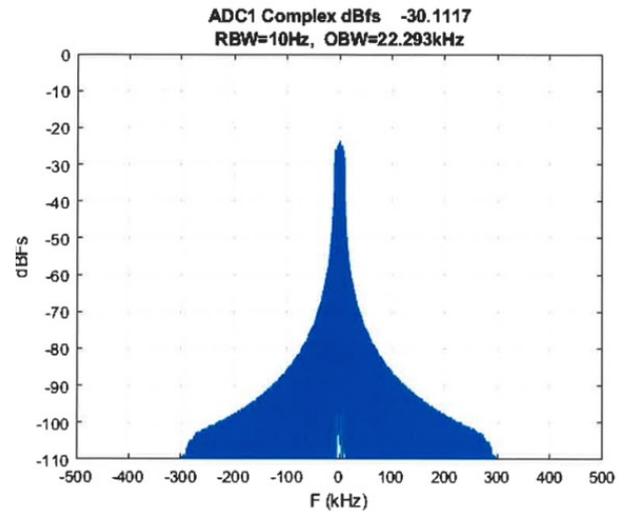

Fig. 4. Costas FSK signal: Transmitter input.

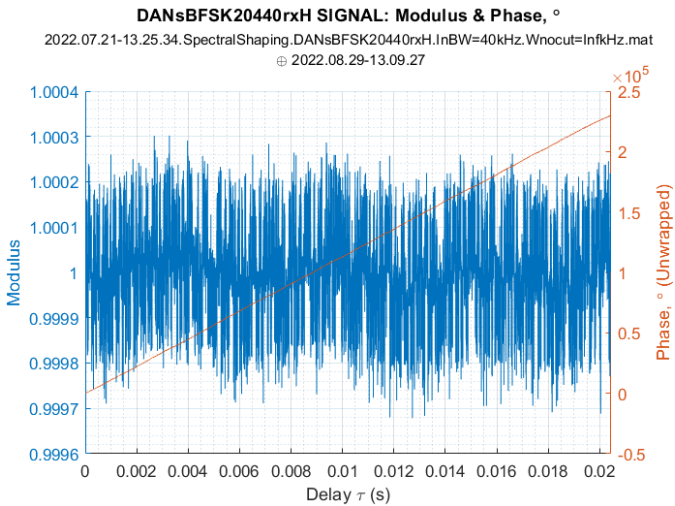

Fig. 3. BFSK signal: Transmitter output modulus and phase.

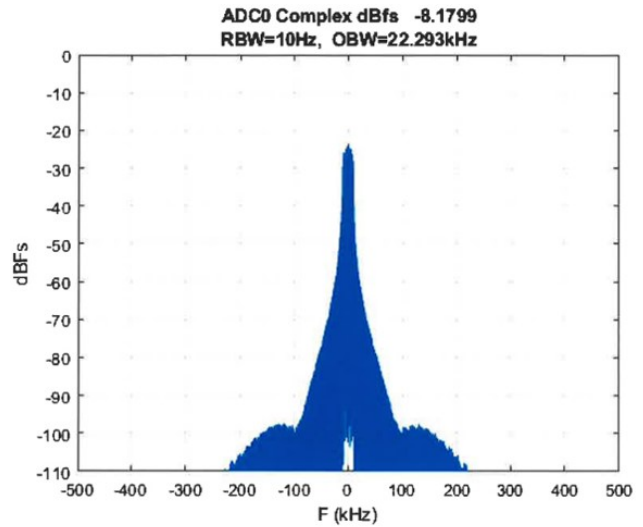

Fig. 5. Costas FSK signal: Transmitter output.

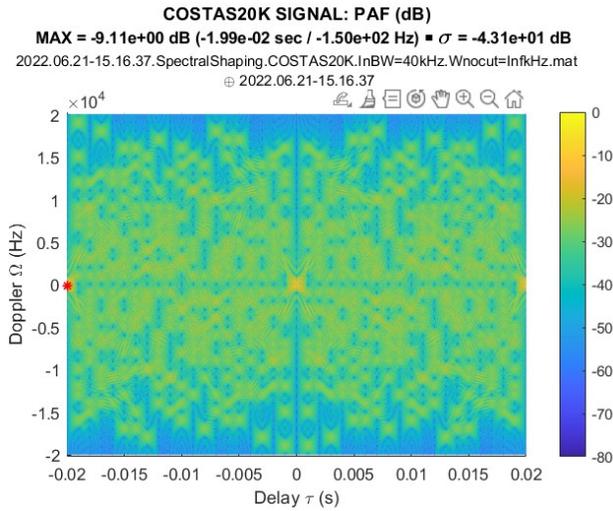

Fig. 6. Original Costas Periodic Ambiguity Function (pAAF)
(max sidelobe = -9.11 dB; σ = -43.1 dB.)

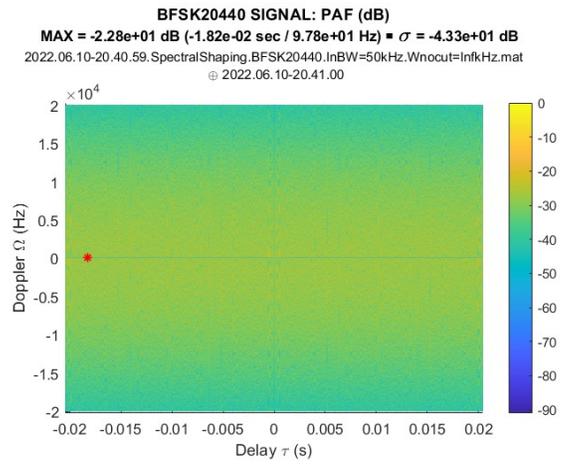

Fig. 8. BFSK 511x40 = 20,440 signal: Ambiguity Function (pAAF).

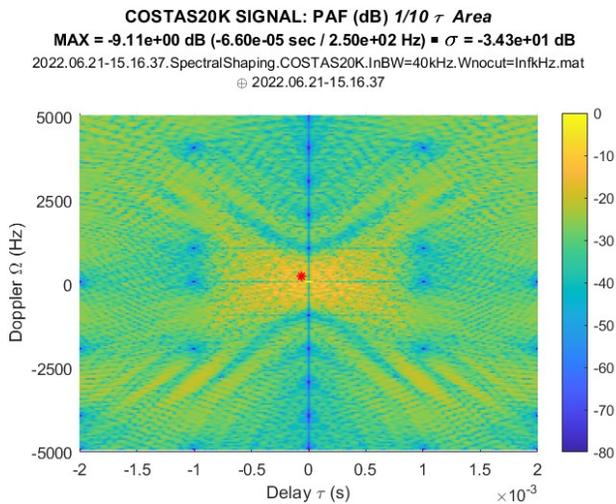

Fig. 7. Original Costas Periodic Ambiguity Function (pAAF) -
the central part (max sidelobe = -9.11 dB; σ = -34.3 dB.)

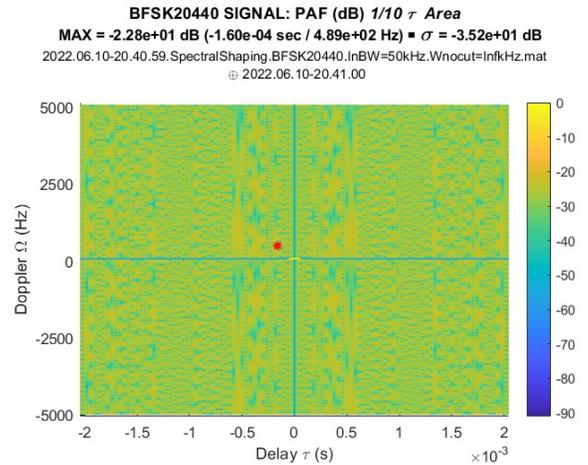

Fig. 9. BFSK 511x40 = 20,440 signal: Ambiguity Function (pAAF) -
the central part (max sidelobe = -22.8 dB; σ = -35.2 dB.)

Indeed, as follows from Fig. 7, there is an area with a consistently high sidelobe level (~10 dB) around the central peak. While this particular Costas sequence is unlikely to be recommended for HF OTHR operations, we included it in our analysis to illustrate a specific phenomenon that spectrum modification can cause.

Finally, in Fig. 10, we compare the spectrum of the ideal LFM CW waveform with a bandwidth of 20 kHz. The spectrum of this waveform at the output of HPA (Fig. 11) remains practically the same. One can observe a relatively modest growth of the occupied bandwidth (5/3 times) with respect to the swept bandwidth of 20 kHz. As discussed in the Introduction, this property is one of those that make this waveform so popular for HF OTHR applications.

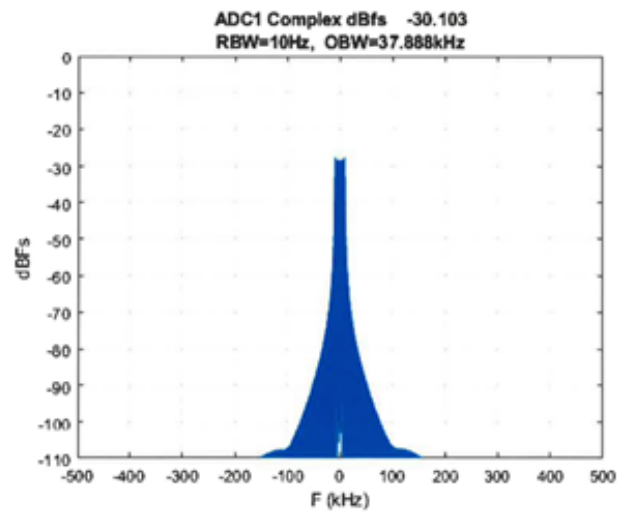

Fig. 10. LFM CW signal: Transmitter input.

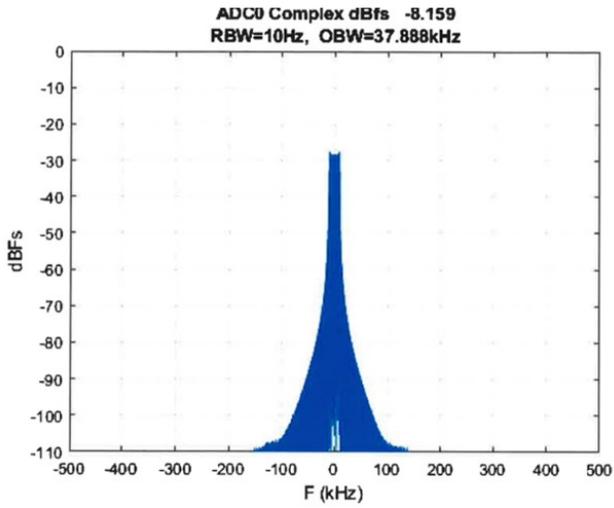

Fig. 11. LFM CW signal: Transmitter output.

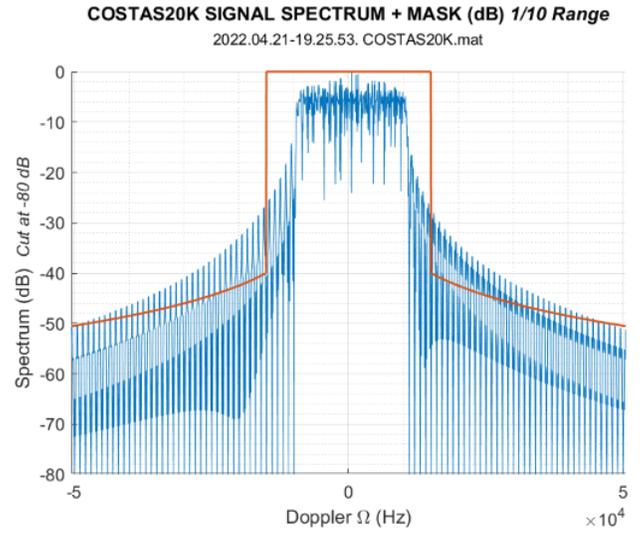

Fig. 12. Original Costas signal:
Spectrum vs. mask (the central part; OBD = 30 dB.)

## III. "AD-HOC" SPECTRAL "TRIMMING" TECHNIQUE

Note that the modulation (deviation) bandwidth of both FSK waveforms is defined by the chip duration $\tau_0$ and the frequency shift $\Delta f$. The actual occupied bandwidth broadening compared with the modulation bandwidth in both these waveforms is associated with abrupt frequency changes.

Therefore, a reasonable ad-hoc approach to reduce this spectrum broadening is to "smooth" the frequency transitions into an analytical frequency modulation function $f_m(t)$. The art within this approach is to find the least "invasive" smoothing filter that, on the one hand, properly "trims" the spectrum of the prototype waveform to meet the NTIA RSEC mask requirements and, on the other hand, to introduce by this smoothing minimal degradation to the ambiguity function of the prototype waveform.

In this example, the original piece-wise frequency modulation function $f_m(t)$ was filtered by a Gaussian filter that "smoothed" the abrupt frequency transitions into an analytic function.

In Fig. 12, the original Costas signal spectrum and the mask are presented.

The modified spectrum and the mask are illustrated in Fig. 13, demonstrating the modified spectrum compliance with the mask for OBD = 30 kHz.

The periodic ambiguity function of the modified Costas waveform is illustrated in Fig. 14 and Fig. 15. One can see that the introduced frequency function "smoothing" did not significantly change the ambiguity function of the prototype Costas waveform (Fig. 6, Fig. 7). Indeed, both the maximum and the MSE sidelobes level and the ambiguity function shape remained practically the same.

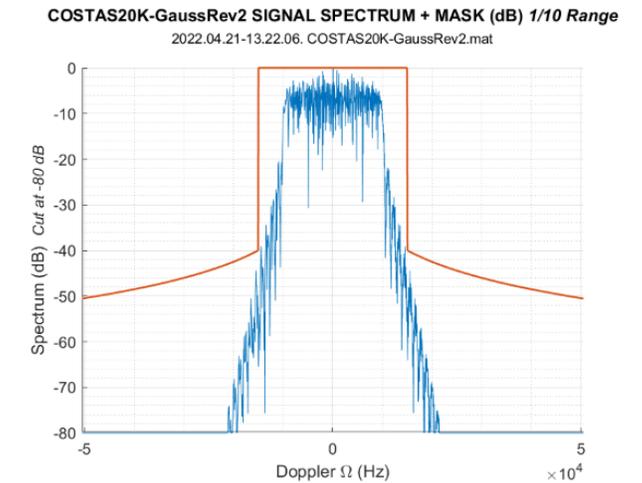

Fig. 13. Spectrum-modified Costas signal:
Spectrum and the mask (the central part; OBD = 30 dB.)

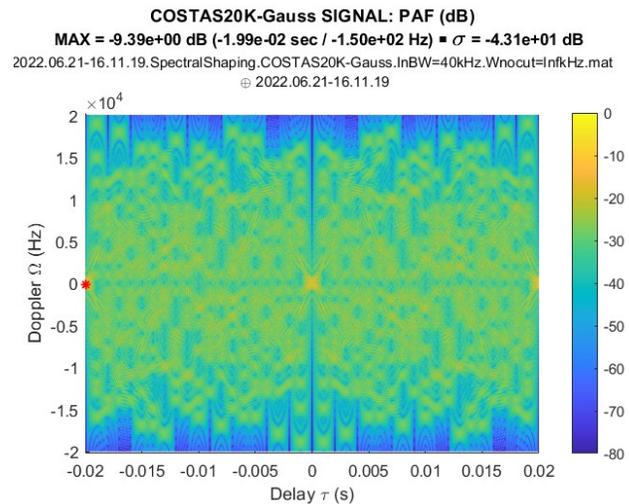

Fig. 14. Modified Costas Ambiguity Function
(OBD = 30 kHz; max sidelobe = -9.39 dB; σ = -43.1 dB.)

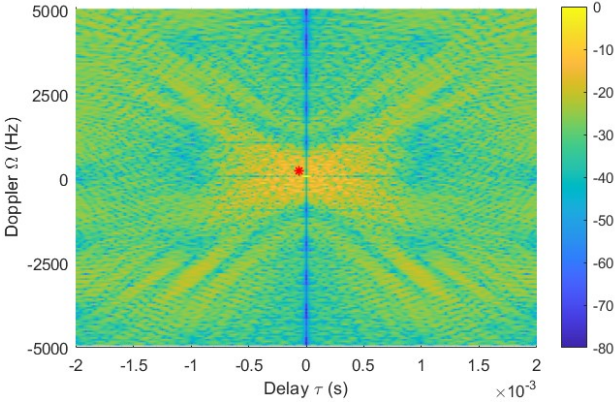

Fig. 15. Modified Costas Ambiguity Function – the central part
(OBD = 30 kHz; max sidelobe = -9.39 dB; σ = -34.2 dB.)

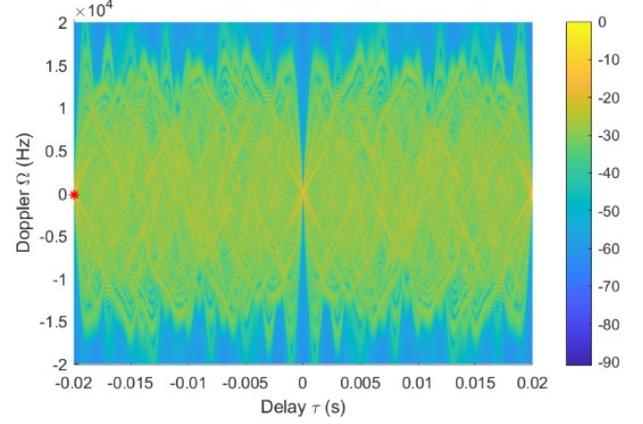

Fig. 17. Spectrum-modified Costas Ambiguity Function
(OBD = 34 kHz; max sidelobe = -11.3 dB; σ = -43.1 dB.)

A more intrusive example of this ad-hoc approach is illustrated in Fig. 16 - Fig. 18, with a more aggressive "smoothing" filtering of the original piece-wise frequency modulation function.

The resulting waveform spectrum (Fig. 16) meets the NTIA RSEC mask requirements for OBD = 34 kHz. Yet, the periodic ambiguity function, illustrated by Fig. 17 - Fig. 18, got two very distinct "ridges" with relatively high sidelobe levels ($\leq 12\ dB \div 13\ dB$) along the two "arms" of the "St. Andrew Cross" formed by these ridges. Note that the maximum and MSE sidelobes level remained practically the same as per the original waveform. It is the new shape with the distinct "arms" that makes the waveform modification as inappropriate as the original waveform or the "slightly" modified previous example.

Therefore, the described above "ad-hoc" technique for "spectral shaping" of the prototype FSK waveforms could be pretty successful. Still, it may require several attempts to find the appropriate filter for the optimal smoothing of the original piece-wise frequency modulation function.

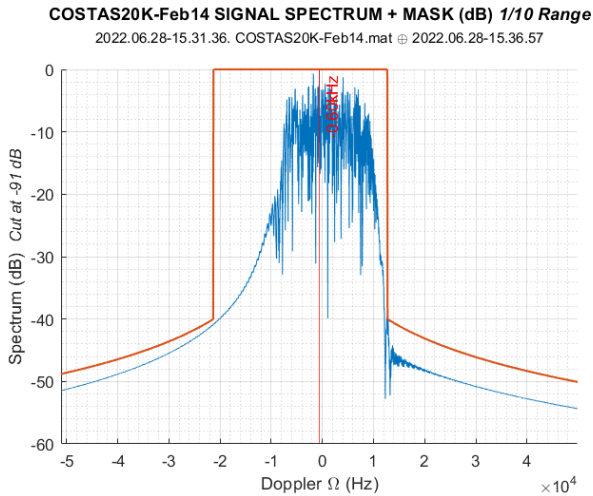

Fig. 16. Spectrum-modified Costas signal
Spectrum vs. mask - the central part (OBD = 34 dB.)

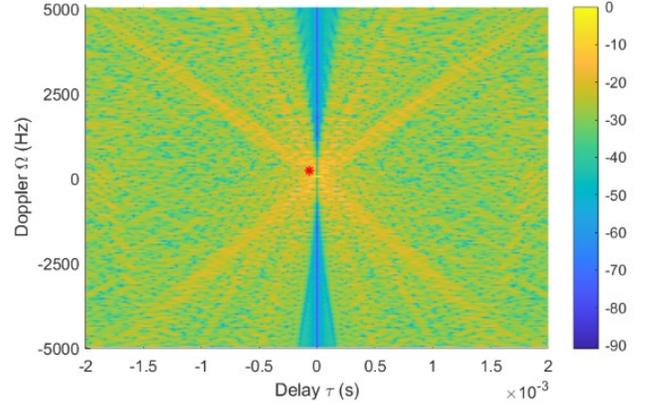

Fig. 18. Spectrum-Modified Costas Ambiguity Function – the central part
(OBD = 34 kHz; max sidelobe = -11.3 dB; σ = -34.1 dB.)

## IV. ALTERNATING PROJECTIONS TECHNIQUE FOR SPECTRUM-CONTROLLED CM WAVEFORM GENERATION

The technique proposed below applies to an arbitrary initial waveform vector, unlike the "ad-hoc" routine described above.

Let $S_0 \in C^{N \times 1}$ be an N-variate vector of the intra-period waveform sampled at a sample rate that exceeds the occupied bandwidth (DBD) of interest.

In our previous examples, the sample rate used in the Digital Waveform Generator (DWG) and digital waveform recorder at the output of the tested radar HPA was $fs = 1$ MHz for all waveforms with OBD that did not exceed 50 kHz.

Let $F_0 \in C^{N \times 1}$ be the N-variate FFT spectrum of the original waveform $S_0$.

The proposed iterative alternating projection technique consists of the following relatively simple steps.

### STEP 1. SPECTRUM $F_0$ "SHAPING"

$$T_1(F_{n-1}) \rightarrow F_n \quad (1)$$

where

$$|F_n| \leq M_B(f) \quad (2)$$

Here, $M_B(f)$ is the NTIA RSEC mask calculated for a given occupied bandwidth OBD = B.

At this step, the spectrum $F_{n-1}$ ($F_0$ for the first iteration with the waveform $S_0$) is modified to fit the mask.

More specifically, for the maximum spectrum values being fixed, all spectrum components whose moduli exceed the mask are replaced by the mask values. The phase content of the spectrum $F_{n-1}$ (or $F_0$) remains unchanged.

The operation is the straightforward projection of the spectrum $F_{n-1}$ onto the closed subspace of the mask-obeying spectra.

### STEP 2. INVERSE FFT (IFFT)

$$IFFT[F_n] \rightarrow S_n \quad (3)$$

Here $S_n \in C^{N \times 1}$ is the waveform vector with the spectrum $F_n$ that meets the mask requirement. In general, this vector $S_n$ has both amplitude and phase modulation.

### STEP 3. PROJECTION OF $S_n$ ON THE CONSTANT MODULUS SET

$$T_2(S_n) \rightarrow S_{n+1} \quad (4)$$

$$|S_{n+1}| = 1, \; arg(S_{n+1})_j = arg(S_n)_j \quad (5)$$

Here we remove amplitude modulation in the waveform $S_n$, leaving its phase modulation intact.

The iterative procedure $(T_2 T_1)^n S_0$ continues until its convergence to a CM waveform with the spectrum that fits NTIA RSEC mask requirements.

It is evident that the projection of the arbitrary waveform spectrum onto the NTIA RSEC mask-fitting spectral subspace, as well as the projection of an arbitrary waveform on the constant modulus waveforms subspace, provide the closest vectors to $S_n$ in the corresponding subspaces and, therefore, are the orthogonal projections. Then, according to the following von Neumann theorem [13]:

**Theorem**. Let $T_1, T_2$ be orthogonal projections onto the closed subspaces $M_1$ and $M_2$ of the real or complex Hilbert space $H$, and $T_M$ the orthogonal projection onto $M = M_1 \cap M_2$.

Then, for any $x \in H$,

$$\|(T_2 T_1)^n x - T_M x\| \rightarrow 0 \text{ as } n \rightarrow \infty \quad (6)$$

our iterative procedure is guaranteed to converge to the norm.

In any practical iterative routine, convergence is defined when the iteration reaches a certain bound from the imposed limit. In our routine, we stop iterations if, after projection onto the CM set, the departure from the imposed spectral mask does not exceed $10^{-3}$ dB in any spectral point.

Naturally, we can stop our iterations after projection onto the spectral mask set, observing some departure from the strict constant modulus property.

Note, in this regard, that spectral limitation by ±250 kHz by the DSP network in the HPA (Fig. 3) resulted in the residual amplitude modulation within the range (-0.9997 ÷ 1.0003) that has no impact on the pACF sidelobe level and the HPA output power.

In most of our "easy" scenarios with an insignificant amount of power to be distributed from the initial solution, we observe the successful convergence of the iterative procedure over a finite number of iterations. Yet, in some "difficult" cases, we limited the total number of iterations by $10^5$ and reported on the results for each of the two final projections.

As discussed above, this routine is applicable for the existing CM waveforms, such as the binary and Costas FSK waveforms, and for an arbitrary realization of a broadband white (Gaussian) noise. Analysis of the application results of this routine to the Costas, binary FSK, and white noise waveform modifications are introduced below.

## IV. A. COSTAS WAVEFORM SPECTRUM-CONTROLLED WAVEFORM MODIFICATION

In this Section, we introduce the results of the iterative alternating projection waveform modification routine (1) – (5) applied to the same Costas waveform used in Section 3 as the prototype for the "ad-hoc" spectral trimming routine.

Unlike the "ad-hoc" routine, here, we can select an arbitrary occupied bandwidth (OBD) that uniquely specifies the NTIA RSEC mask.

For DBD = 20 kHz, we selected OBD = 20 kHz, 25 kHz, 30 kHz, 35 kHz, and 40 kHz. For each of the examples mentioned above, in the figures below, we illustrate the following:

**Convergence**
As the metric of the convergence of our iterative procedure, we use the sum of the spectral components (in dB) exceeding the mask border after projection onto the CM subspace (Fig. 19).

**Ambiguity function sidelobes level**
It is provided for the original Costas waveform and spectrum-modified versions (Table I).

For the most challenging case with OBD = DBD = 20 kHz, we illustrate the modified waveform spectrum (Fig. 20) and the periodic ambiguity function (Fig. 21 and Fig. 22).

Comparison with the original periodic ambiguity function (Fig. 6 - Fig. 7) demonstrates practically identical ambiguity functions of the original and spectrum-modified Costas waveforms.

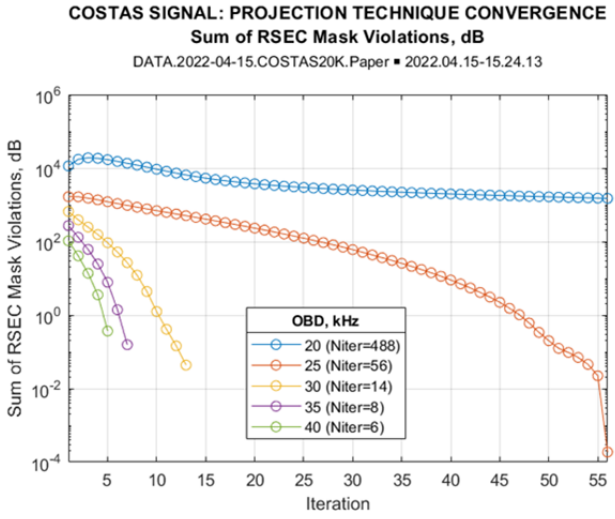

Fig. 19. Spectrum-modified Costas signal:
Projection technique conversion rate.

**TABLE I**
**Spectrum-Modified Costas Signal:**
**Projection Technique pAAF Metrics**

| AREA | pAAF Metrics | Original Signal | OBD, kHz | | | | |
|---|---|---|---|---|---|---|---|
| | | | 20 | 25 | 30 | 35 | 40 |
| FULL | MAX Sidelobe, dB | -9.30 | -10.39 | -9.35 | -9.32 | -9.31 | -9.30 |
| | $\sigma$, dB | -35.16 | -35.18 | -35.10 | -35.13 | -35.14 | -35.15 |
| CENTRAL | MAX Sidelobe, dB | -9.30 | -10.39 | -9.35 | -9.32 | -9.31 | -9.30 |
| | $\sigma$, dB | -28.09 | -28.41 | -28.12 | -28.09 | -28.09 | -28.09 |

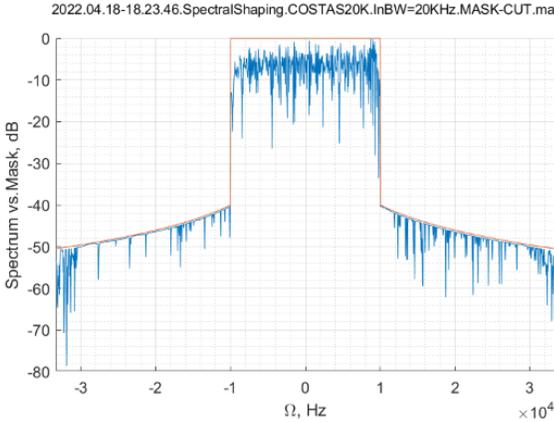

Fig. 20. Spectrum-modified Costas signal
The final spectrum vs. mask - the central part (OBD = 20 kHz.)

As the introduced results show, the convergence rate is slower for a more stringent spectrum restriction than the original waveform spectrum, i.e., for a narrower OBD. The longest convergence (488 iterations) we observed for OBD = BDB = 20 kHz.

The fastest convergence rate (6 iterations) we observed for OBD = 40 kHz.

This result is quite expected since, for more stringent spectrum restrictions, a more significant portion of the prototype waveform spectrum should be re-distributed.

A comparison of the sidelobes levels in Table I of the original Costas waveform with the maximum and mean square of the sidelobes level of the final spectrum-controlled waveform demonstrates insignificant changes in the sidelobes level values caused by spectrum modification.

Indeed, the maximal sidelobe level observed in the close vicinity of the central peak equal to – 9.30 dB remained practically the same (-10.39 dB) even for the most stringent mask with OBD = 20 kHz.

For OBD > 25 kHz, these distinctions are even less pronounced.

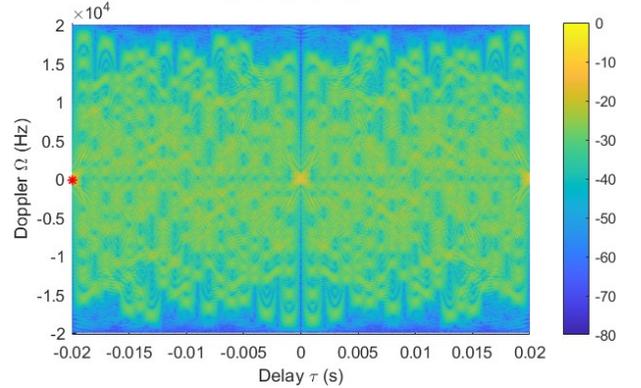

Fig. 21. Optimized Costas signal: Periodic Ambiguity Function (pAAF)
(OBD = 20 kHz; max sidelobe = -9.74 dB; $\sigma$ = -43.1 dB.)

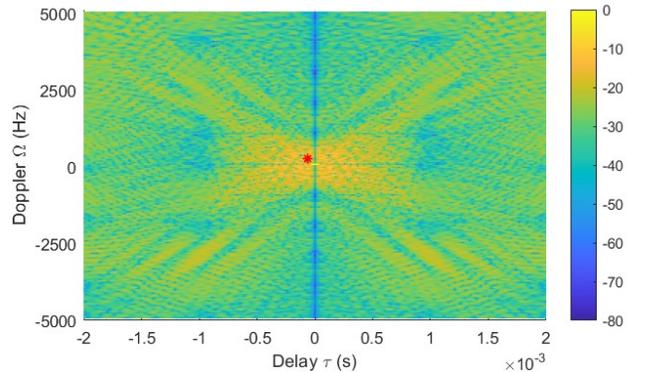

Fig. 22. Optimized Costas Signal
Periodic Ambiguity Function (pAAF) – the central part
(OBD = 20 kHz; max sidelobe = -9.74 dB; $\sigma$ = -34.2 dB.)

Yet, as discussed above, the particular Costas analyzed waveform is not the best example of a truly "thumb-tack" ambiguity function property due to the high sidelobes level in the close vicinity of the ambiguity function peak.

Therefore, analysis of the waveform with a very low sidelobes level of the periodic ACF function of the ideal BFSK waveform should be much more illustrative.

## IV. B. BINARY FSK WAVEFORM SPECTRUM-CONTROLLED MODIFICATION

For this analysis, as a prototype, we use the N = 511-element BFSK waveform with the modulation (deviation) bandwidth DBD = $2\Delta f$ = 25.0 kHz, $\Delta f = 1/2\tau_0$ = 12.5 kHz when WRF = 48.9237 Hz and $N$ = 20,440.

This number of samples ($N = 40 \times 511$) is easy to downsample to any number of samples, multiple of 511 ($N$ = k×511, k = 1, 2, …, 40).

First of all, we demonstrate that, indeed, the pACF of this waveform, even downsampled to N = 511, has a sidelobe level of -54.17 dB (Fig. 23).

Now, we can apply our alternating projection spectrum modification routine, mainly focusing on the periodic ACF sidelobe level.

The very low sidelobe level of the original pACF (-54.17 dB) provides an opportunity to investigate the dependence of this level on the waveform spectrum modification.

First, in Fig. 24, we introduce the unmodified spectrum with the NTIA RSEC mask OBD, which preserves this spectrum. This figure shows that the occupied bandwidth must reach 130 kHz to retain this waveform unmodified.

It is evident that for the waveform with a deviation bandwidth of 25.049 kHz, securing such broad bandwidth for HF OTHR operations is impractical.

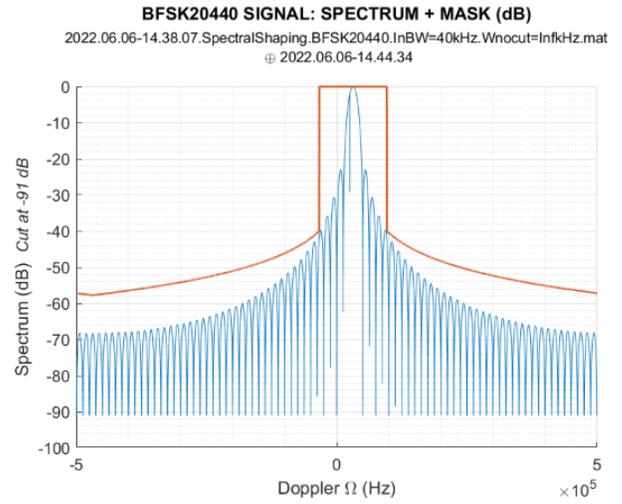

Fig. 24. Original BFSK 20,440 signal: Spectrum vs. mask
(OBD = 130 kHz.)

To address this question, we modified the original waveform spectrum to reach -90 dB beyond a specified bandwidth. For this analysis, we had to limit, in most cases, the total number of iterations by $N = 10^5$ and report on the results of the two final projections.

We start from the projection onto the family with the prescribed spectrum.

Fig. 25 presents the spectrum with the Full Bandwidth FBD = 100 kHz. As one can see, before being driven to the -90 dB level, the "powerful" spectrum reached -50 dB, while in the original spectrum (Fig. 24), it is still well above -40 dB at these "border" frequencies.

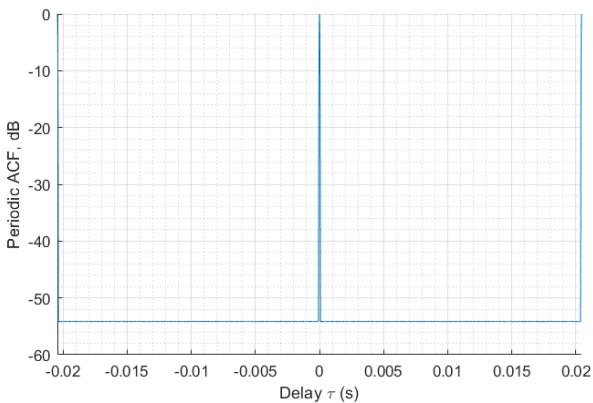

Fig. 23. BFSK 20,440 signal downsampled to N = 511
Periodic ACF (pACF); max sidelobe = -54.17 dB

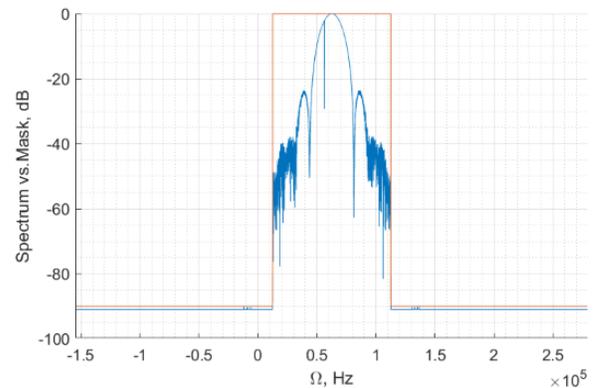

Fig. 25. Spectrum-modified BFSK 20,440 final signal:
Spectrum vs. mask
(OBD = 130 kHz; FBD = 100 kHz; Niter = 17,358.)

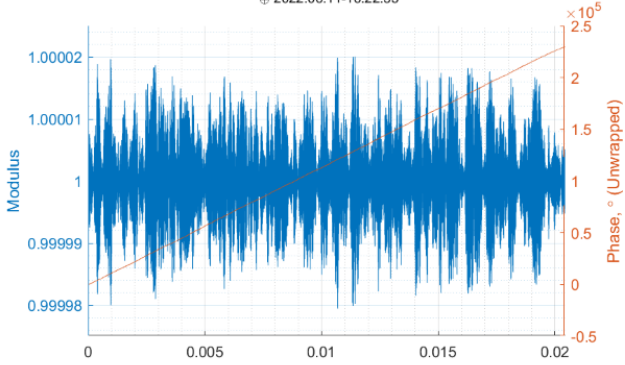

Fig. 26. Spectrum-Modified BFSK 20,440 "Raw" Signal:
Residual amplitude modulation
(OBD = 130 kHz; FBD = 100 kHz; Niter = 17,358.)

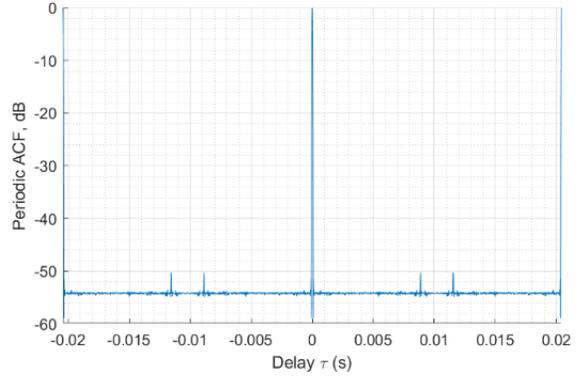

Fig. 28. Spectrum-modified BFSK 20,440 final signal
Periodic ACF (pACF)
(MAX Sidelobe = -50.22 dB; FBD = 100 kHz; Niter = 17,358.)

Note that our formal convergence criterion imposes bounds on the deviation from the spectral mask for a strictly CM waveform. Yet, as expected, severe spectrum truncation to the mask results in the residual amplitude modulation, illustrated in Fig. 26 (0.99998 – 1.00002.) This modulation is similar to the one observed in Fig. 3 and therefore does not affect the output HPA power.

The thumb-tack nature of the pAAF is illustrated in Fig. 27.

The pACF of this optimized waveform with the truncated 100 kHz spectrum is illustrated in Fig. 28, demonstrating the appearance of four range sidelobes reaching -50 dB.

Therefore, by transitioning from the original spectrum with OBD = 130 kHz to the total bandwidth of 100 kHz, we lost ~ 4 dB in the sidelobe level.

In terms of convergence, modification of the original BFSK waveform spectrum to the waveform with the "chimney-like" spectrum with total bandwidth (FBD = 100 kHz) was successful since spectrum convergence of the CM waveform to the mask was achieved at the 17,358$^{th}$ iteration.

For the following example with the FBD = 60 kHz, such convergence was not achieved over 100,000 iterations, and therefore we report on both projections results, completed at the final 100,000$^{th}$ iteration.

Fig. 29 illustrates the ideal "chimney-like" spectrum with the FBD = 60 kHz. Naturally, strict spectrum projection resulted in some remaining amplitude modulation, shown in Fig. 30.

On the contrary, the final projection onto the CM family leads to a minor violation of the spectrum. Indeed, in Fig. 31, beyond 60 kHz bandwidth, one can see two ~ 30 kHz wide "humps" reaching the level of -80 dB, which is 10 dB above the -90 dB mask level.

Yet, the sidelobe level of the pACF for the strictly CM waveform (Fig. 32) and for the waveform that precisely meets the spectral mask (Fig. 33) are almost the same. In both cases, we observe four groups of sidelobes, significantly exceeding the ~50 dB sidelobes "floor" and reaching -33.7 dB.

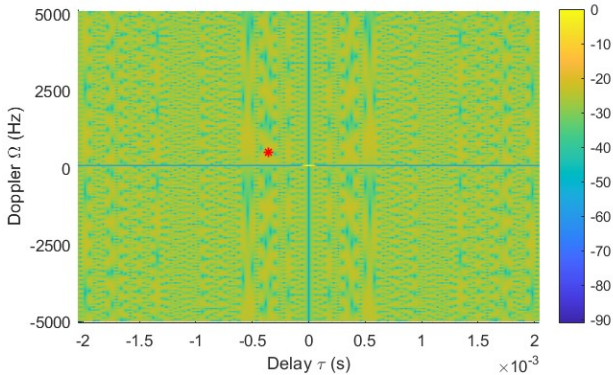

Fig. 27. Spectrum-modified BFSK 20,440 Final signal
Ambiguity Function: max sidelobe = -22.8 dB; σ = -35.1 dB
(OBD = 130 kHz; FBD = 100 kHz.)

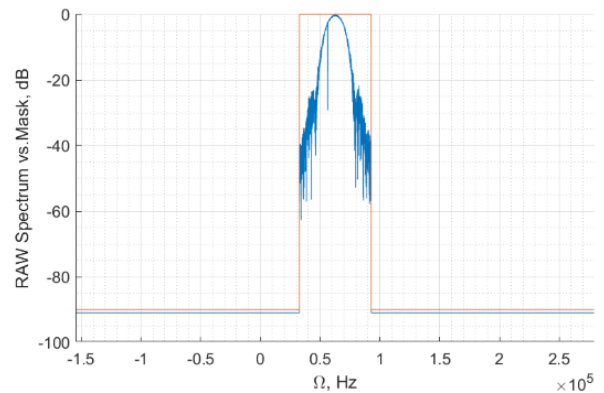

Fig. 29. Spectrum-modified BFSK 20,440 "raw" signal:
Spectrum vs. mask
(OBD = 130 kHz; FBD = 60 kHz; Niter = 100,000.)

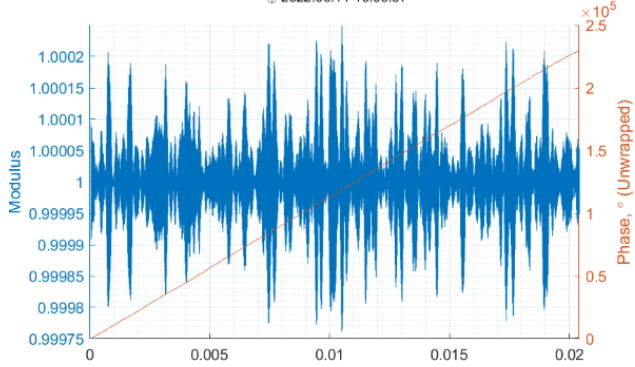

Fig. 30. Spectrum-modified BFSK 20,440 "raw" signal:
Residual amplitude modulation
(OBD = 130 kHz; FBD = 60 kHz; Niter = 100,000.)

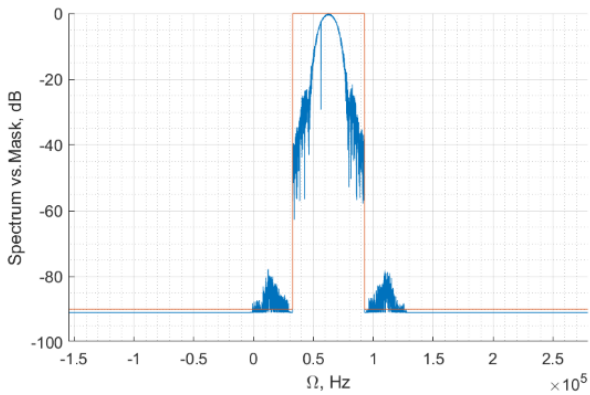

Fig. 31. Spectrum-modified BFSK 20,440 final signal:
Spectrum vs. mask
(OBD = 130 kHz; FBD = 60 kHz; Niter = 100,000.)

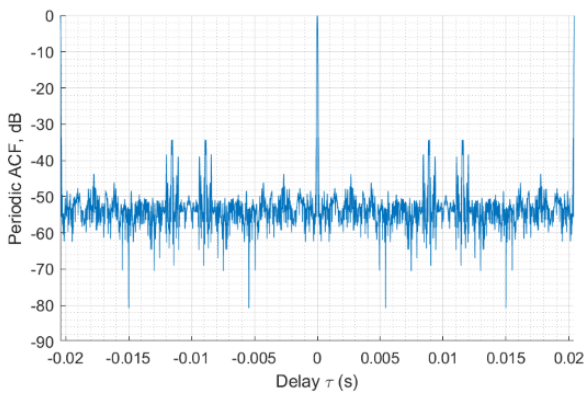

Fig. 32. Spectrum-Modified BFSK 20,440 Final Signal
Periodic ACF (pACF): max sidelobe = -33.7 dB
(OBD = 130 kHz; FBD = 60 kHz; Niter = 100,000.)

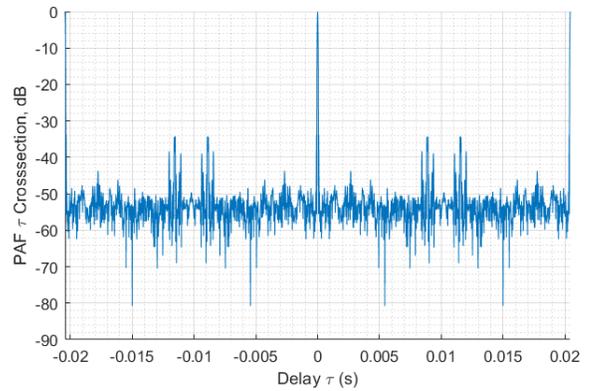

Fig. 33. Spectrum-modified BFSK 20,440 "raw" signal
Periodic ACF (pACF): max sidelobe = -33.7 dB
(OBD = 130 kHz; FBD = 60 kHz; Niter = 100,000.)

The retained thumb-tack nature of the pAAF is illustrated in Fig. 34.

Therefore, even though formal convergence was not achieved in this case, both final iterations of the iterative routine (3) resulted in practically identical pAAF (and pACF.)

Note that amplitude modulation of the spectrally strictly limited by the mask waveform (Fig. 30) within the range (0.9998 ÷ 1,0002) is similar to the "natural" residual amplitude modulation in Fig. 3, which does not affect the HPA output power.

In Table II, we present the results of this iterative routine with the final projection onto the spectral mask family for different bandwidths of the spectral "chimney." We introduce here the maximum pACF sidelobe level, the maximum ambiguity function (pAAF) sidelobe level, the amplitude modulation depth for the final projection of the meeting spectral mask family, and the maximum violation of the -90 dB mask level (in dB), for the final projection onto the CM family.

One can see that we have not achieved the strict convergence point due to the minimal remaining violations of the spectral mask and the CM properties. Yet, the results of both final projections are practically the same.

Another manifestation of the complex interconnection between the pACF sidelobe level and convergence conditions is observed for a 60 kHz sidelobe level of -33.7 dB, which is slightly worse than the sidelobe level of -35.5 dB we got for the restricted bandwidth of 50 kHz.

While violations of the mask and constant modulus conditions on the final 100,000th iteration are smaller for BD = 60 kHz, the 2 dB loss in the sidelobe level compared with the 50 kHz case may be partly due to the unachieved strict convergence conditions and partly since the proposed iterative routine does not guarantee that for the given bandwidth and CM restrictions, we should achieve the minimum possible sidelobe level.

**TABLE II**
**BFSK 20,440 Signal: Iterative Projection Routine Results**
for OBD = 130 kHz; Niter = 100,000

| # | FBD, kHz | Step 1: Projection on Spectrum Mask (MASK Violation = 0) | | | Step 2: Projection on CM (CM Violation = 0) | | |
|---|---|---|---|---|---|---|---|
| | | MAX CM Violation | MAX pACF Sidelobe, dB | MAX pAMF Sidelobe, dB | MAX Mask Violation | MAX pACF Sidelobe, dB | MAX pAMF Sidelobe, dB |
| 1 | 60 | 2.489E-04 | -3.368E+01 | -2.146E+01 | 1.223E+01 | -3.368E+01 | -2.146E+01 |
| 2 | 50 | 3.488E-04 | -3.545E+01 | -2.082E+01 | 1.605E+01 | -3.545E+01 | -2.082E+01 |
| 3 | 40 | 4.472E-04 | -2.588E+01 | -1.940E+01 | 1.573E+01 | -2.588E+01 | -1.940E+01 |
| 4 | 30 | 4.233E-04 | -1.470E+01 | -1.470E+01 | 1.423E+01 | -1.470E+01 | -1.470E+01 |
| 5 | 20 | 8.196E-04 | -9.727E+00 | -9.727E+00 | 1.416E+01 | -9.727E+00 | -9.727E+00 |

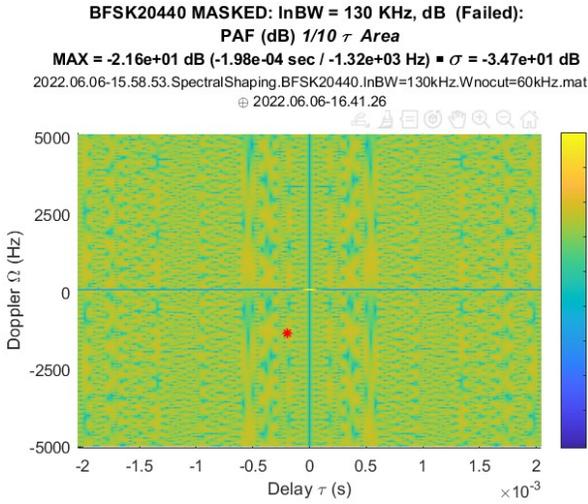

Fig. 34. Spectrum-modified BFSK 20,440 final signal Ambiguity Function: OBD = 130 kHz; FBD = 60 kHz (Max sidelobe = -22.8 dB; σ = -35.1 dB.)

Even though our original BFSK waveform has a much lower sidelobe level of -54.17 dB, there is no guarantee that, treated as the initial solution for our iterative procedure, it should provide the minimal pACF sidelobe level for any strictly restricted bandwidth. In this regard, the problem of CW CM waveform synthesis with a strictly limited bandwidth remains an open problem. At the same time, from the practical standpoint, the 2 dB loss in the sidelobe level is not that critical.

One can also see that the waveform spectrum truncation at the bandwidth of 30 kHz results in the -14.7 dB sidelobe level, and only for the FBD = 40 kHz do we get the noise-like sidelobe level $(\sqrt{N})$ where the waveform bandwidth must significantly exceed the modulation bandwidth of 25 kHz.

Note that above we performed spectral modifications that exceed the NTIA requirements but suit the practical conditions of the HF OTHR operations.

Still, in Table III, we introduce the results of our iterative routine to meet the NTIA RSEC criterion mask requirements for the different OBDs within the range of 20 kHz ÷ 60 kHz.

As one can see, the low-level part of the waveform spectrum, retained during these iterations, allowed for keeping a significantly lower pACF sidelobe level.

Indeed, the "chimney-like" restriction of the entire spectrum by 60 kHz resulted in a pACF sidelobe level of -33.7 dB (Table I) while meeting the NTIA mask with OBD = 60 kHz, retained the sidelobes level at the level below -49.8 dB (Table II).

Therefore, retaining the spectral components below the -40 dB level improved the sidelobe level by 15 dB.

The most important observation that follows from this analysis is that to retain the pACF sidelobe level below the "noise-like" limit $\sqrt{N}$, the spectrum of the waveform should significantly exceed the modulation (deviation) bandwidth.

This "noise-like" CM continuous-wave waveforms synthesis is introduced in Section IV.C.

## IV. C. CONSTANT MODULUS NOISE-LIKE SPECTRUM-CONTROLLED WAVEFORM GENERATION

The observed noise waveform-like pACF sidelobe level we achieved when the modified BFSK waveform spectrum bandwidth truncated close to its modulation bandwidth suggests that other waveforms with a worse original sidelobes level may be used as the initial ones in our iterative routine.

Particularly, since for our alternating projection routine, the initial waveform may be arbitrary, i.e., with amplitude and phase modulations, we may use a realization of the broadband noise as the initial waveform.

TABLE III
BFSK 20,440 Signal: Iterative Projection Routine Results For FBD = ∞

| # | OBD, kHz | # of Iter. | Step 1: Projection on Spectrum Mask (MASK Violation = 0) | | | Step 2: Projection on CM (CM Violation = 0) | | |
|---|---|---|---|---|---|---|---|---|
| | | | MAX CM Violation | MAX pACF Sidelobe, dB | MAX pAMF Sidelobe, dB | MAX Mask Violation | MAX pACF Sidelobe, dB | MAX pAMF Sidelobe, dB |
| 1 | 60 | 6 | 2.642E-03 | -4.987E+01 | -2.282E+01 | 0.000E+00 | -4.979E+01 | -2.282E+01 |
| 2 | 50 | 53 | 2.751E-03 | -3.746E+01 | -2.245E+01 | 0.000E+00 | -3.744E+01 | -2.245E+01 |
| 3 | 40 | 94 | 4.286E-03 | -3.186E+01 | -2.131E+01 | 0.000E+00 | -3.185E+01 | -2.131E+01 |
| 4 | 30 | 281 | 3.150E-03 | -2.796E+01 | -2.001E+01 | 0.000E+00 | -2.796E+01 | -2.001E+01 |
| 5 | 20 | 244 | 3.611E-03 | -1.385E+01 | -1.385E+01 | 0.000E+00 | -1.385E+01 | -1.385E+01 |

To demonstrate this methodology, we treated the sequence of $20\times10^3$ i.i.d. complex Gaussian random numbers as the initial waveform. This sequence may be treated as the samples with the sampling rate of 1 MHz white noise realization over the repetition period with WRF = 50 Hz.

Apparently, the ambiguity function (**Error! Reference source not found.**) of this noise waveform is thumb-tack, with the mean square sidelobe level close to $\sqrt{M}$ = 43 dB.

In Fig. 36, we present the pACF, while in Fig. 37, we illustrate close to the constant spectrum of the waveform with a 1 MHz-wide band and the spectral mask the modified CM waveform has to obey. For the OBD = 20 kHz, we allowed the total bandwidth (above -90 dB level) to be only 40 kHz wide.

Our iterative routine converged after 95,656 iterations, meeting $10^{-3}$ dB margins for the spectrum mask (Fig. 38), achieved after amplitude fluctuations truncation.

Strict spectrum truncation to the -90 dB mask level resulted in an insignificant amplitude modulation within the margins (0.99996 ÷ 1.00006) that does not affect the pAAF sidelobe level or the transmitted power.

The mean square sidelobe level of the pAAF is the same as for the entire pAAF (-43.1 dB, Fig. 39), while the maximum pACF sidelobe level reached -14.7 dB (Fig. 40.)

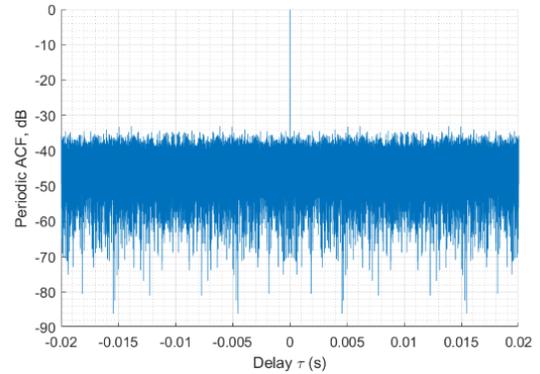

Fig. 36. Random 20K signal: Periodic ACF (pACF)
(Max sidelobe = -33.0 dB.)

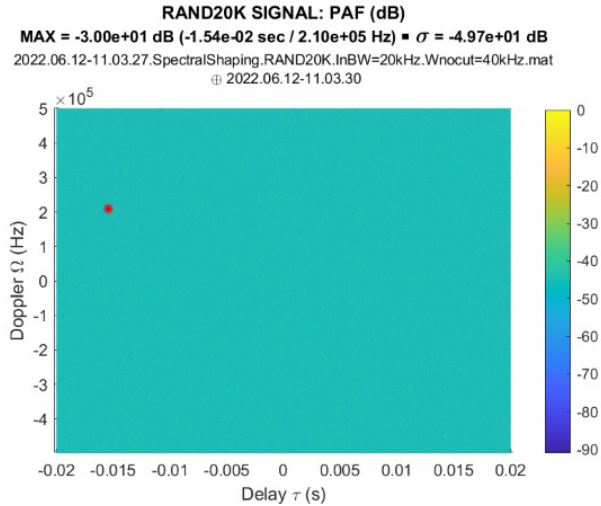

Fig. 35. Random 20K signal: Ambiguity Function (pAAF)
(Max sidelobe = -30.0 dB; σ = -49.7 dB.)

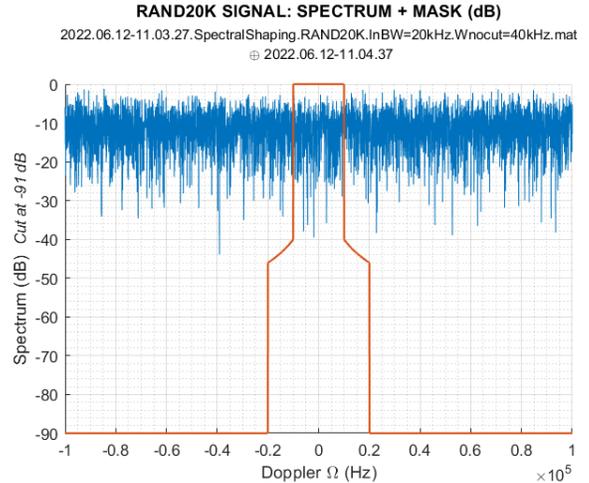

Fig. 37. Random 20K signal: Spectrum vs. Mask
(OBD = 20 kHz; FBD = 40 kHz.)

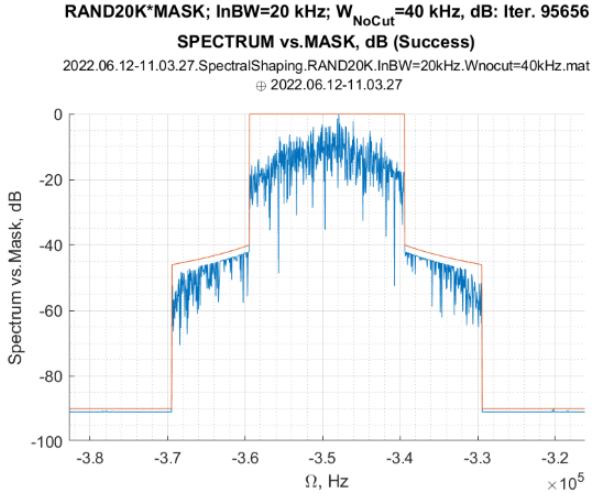

Fig. 38. Final optimized random 20K signal:
Spectrum vs. mask
(OBD = 20 kHz; FBD = 40 kHz; Niter = 95,656.)

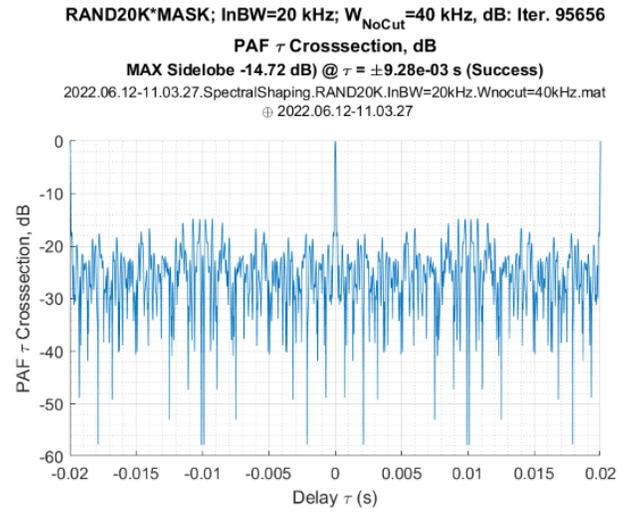

Fig. 40. Final optimized random 20K signal
Periodic ACF (pACF): Max Sidelobe = -14.7 dB
(OBD = 20 kHz; FBD = 40 kHz; Niter = 95,656.)

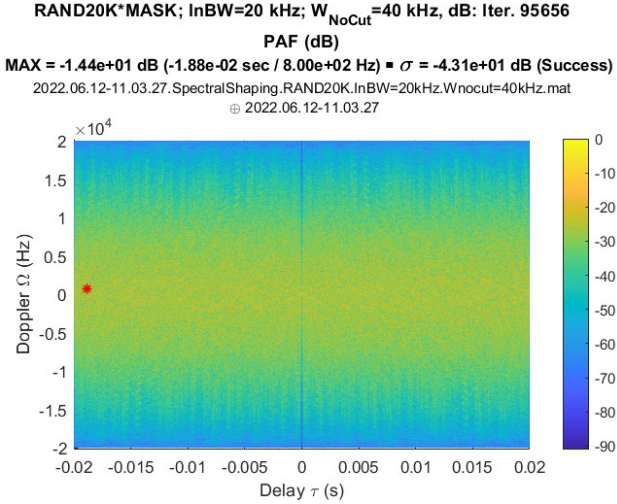

Fig. 39. Final optimized random 20K signal
Periodic Ambiguity Function (pAAF)
(Max Sidelobe = -14.4 dB; σ = -43.1 dB.)

The introduced example demonstrates our ability to generate a constant-modulus (CM) noise-like waveform with the thumb-tack ambiguity function and very high spectral occupancy.

Note that a similar sidelobe level should be observed for the cross-ambiguity function of the similarly generated waveforms that used independent noise realizations to initiate the alternating projection routine. This set of waveforms may be used for MIMO applications [12] and in other applications where waveform diversity is essential.

The main drawback of the generated CW CM waveforms is the relatively high range sidelobe level associated with the "chimney-like" spectrum shape.

Note in this regard that the specifics of HF OTHR operations dictate the "chimney-like" spectrum shape, which is a more stringent restriction than the one required by the NTIA RSEC Criterion C.

These radars operate in an unlicensed spectrum. Therefore, if broad bandwidth occupancy is required by the waveform, the user faces a difficult proposition to locate such spectral real estate and hold on to it for extended periods of time. This is especially true for signals that have large swaths of very low power spectral density, as these will likely appear to another unlicensed user as an unoccupied spectrum.

Indeed, some HF users may treat the low-power sub-bands of the waveform spectrum as free and occupy them.

To avoid this, all spectral components of the transmitted waveform should be sufficiently high power to be easily identified by HF users. The relatively high pACF sidelobe level limits the dynamic range of jointly detected targets that share the same azimuth and Doppler frequency.

To expand this dynamic range and detect weak targets masked by the range sidelobes of the strong ones, different mismatched processing techniques may be applied.

These techniques are introduced and analyzed in Part II of this series.

## V. CONCLUSIONS AND RECOMMENDATIONS

In this paper, we demonstrated that the known CM waveforms with the thumb-tack ambiguity function occupy a bandwidth specified by the NTIA RSEC methodology that significantly exceeds these waveforms' modulation (deviation) bandwidth.

This occupied bandwidth broadening is due to the abrupt phase or frequency changes in phase- or frequency-modulated waveforms correspondingly.

For HF OTH radars, this broadening is prohibitive since it requires a search, allocation, and "kept custody" of the correspondingly broad frequency channels unoccupied by other HF users.

In this study, we introduced an "ad-hoc" technique for "smoothing" the abrupt frequency changes in the frequency shift keying (FSK) waveforms, such as BFSK [6, 7] and Costas [9] waveforms. A universal alternating projection routine was proposed to convert any broadband sequence into a (periodic) constant modulus (CM) waveform, accurately meeting the NTIA RSEC Criterion C mask.

All these waveforms retain the thumb-tack property of their original periodic ambiguity function after spectrum modification, though some essential properties may be affected by spectrum modification.

In particular, we demonstrated that the remarkable property of the BFSK periodic autocorrelation function with the sidelobe level equal to -20lgN, N ~ OBD/WRF, is susceptible to spectrum modification: any attempt to restrict its bandwidth leads to the rapid sidelobe level degradation.

On the other hand, the noise-like mean square sidelobe level of the entire periodic ambiguity function (~10lgN) is quite robust concerning the spectrum modification.

We demonstrated that the proposed alternating projection technique might be initiated by an arbitrary (broadband) vector, not necessarily representing a known CM waveform.

In particular, realizations of the broadband white (Gaussian) noise were used for the initialization of our routine, resulting in a family of constant modulus waveforms that accurately fit the NTIA RSEC mask with a prescribed occupied bandwidth (OBD) and the mean square level of auto - and cross-ambiguity function sidelobes equal to ~ -10log [OBD/WRF].

We demonstrated that this methodology could be used to generate the set of CM waveforms that not only meet NTIA RSEC mask requirements but have the required spectrum efficiency with the entire spectrum concentrated within the prescribed bandwidth.

This paper's spectrum-controlled constant modulus waveform generation methodology focuses on an application in HF OTHR radars. Still, it may be useful for microwave radars recently sharing frequency spectrum with mobile communication networks [10, 11].

High spectrum efficiency of the generated CW CM waveforms with the thumb-tack ambiguity function is associated with the relatively high range sidelobes with the mean square level of (-10 log [OBD/WRF]) dB.

This sidelobes level limits the dynamic range of jointly detected targets with the same angle of arrival and Doppler frequency if matched range processing is used.

To increase this dynamic range and facilitate weak target detection masked by range sidelobes of the strong ones, several mismatched range processing techniques may be used.

When applied to these waveforms, the efficiency of these techniques in Direct Digital Receivers of the modern HF OTHR is analyzed in Part II of this series.

# REFERENCES

[1] D. H. Sinnott, "Over the Horizon Radar Down-Under," in Proc. of the 2015 IEEE Radar Conference (RadarCon), VA, USA, 2015, pp. 1761–1764.

[2] Royal Australian Air Force Fact Sheet," Jindalee Operational Radar Network." [online] Available: https://www.airforce.gov.au/docs/JORN_Fact_Sheet_.pdf

[3] J. M. Hedrick, J. F. Thomason, "Naval Applications of the High-Frequency Over-the-Horizon Radar," Naval Engineers Journal, March 2009, 108 (3), pp. 353-362, [online] Available: https://trid.trb.org/view.aspx?id=480808#:~:text=NAVAL%20APPLICATIONS%20OF%20HIGH%20FREQUENCY%20OVER-THE-HORIZON%20RADAR%20The,area%20and%20out%20to%20ranges%20of%202000%20nmi.

[4] M.P. Harnett, J.T. Clancy, R. J. Denton, "Utilization of a Non-recurrent Waveform to Mitigate Range-Folded Spread Clutter: Application to Over-The-Horizon Radar," Radio Science, vol. 33, no. 4, pp. 1125-1133, Jul.-Aug. 1998.

[5] Manual of Regulations and Procedures for Federal Radio Frequency Management, National Telecommunications and Information Administration (NTIA), US Department of Commerce, 2021 Edition, Jan. 2021.

[6] N. Levanon, I. Cohen, "Binary Frequency Shift Keying for Continuous Wave Radar," IEEE Trans. AES, vol. 53, no. 3, Oct. 2017, pp. 2462-2468.

[7 N. N. Levanon, Radar—Concise Course Volume 2, [online] Available: http://www.eng.tau.ac.il/~nadav/levanon_radar_course_vol_2.pdf.

[8] G.F.Earl, B.D. Ward, "The Frequency Management System of the Jindalee Over-the-Horizon Backscatter HF Radar," Radio Science, vol. 22, no. 2, March-April 1987, pp. 275-291.

[9] B. Correll, J.K. Beard, C.N. Swanson, "Costas Array Waveforms for Closely Spaced Target Detection," IEEE Trans. on AES, vol.56, No.2, Apr. 2020, pp. 1045-1076.

[10] Defense Advanced Research Project Agency (DARPA)," Shared Spectrum Access for Radar and Communications, (SSPARC), DARPA BAA -13-24, 2013, (online) Available: www.darpa.mil.

[11] H. Hayvaci, B. Talvi, "Spectrum Sharing in Radar and Communication Systems: A review," in Proc. Int. Conf. IEEE Electromagn. Adv. Appl. (ICEAA), Aug. 2014, pp. 810-811.

[12] Y.I. Abramovich, G.J. Frazer, B. A. Johnson, "Principles of Mode-Selective MIMO OTHR," IEEE Trans. on AES, vol.49, No. 3, Jul. 2013, pp. 1839-1868.

[13] J. Von Neumann, "On Rings of Operators. Reduction Theory". Annals of Mathematics, 1949, pp. 401-485.